\documentstyle[11pt]{article}
\input epsf

\newcommand{\om}{\omega}

\newcommand{\pa}{\partial}
\newcommand{\be}{\begin{equation}}

\newcommand{\ee}{\end{equation}}
\newcommand{\bea}{\begin{eqnarray}}
\newcommand{\eea}{\end{eqnarray}}
\begin{document}

\begin{flushright}
UCY-PHY-95/8\\
PSI-PR-95-12
\end{flushright}

\begin{center}
{\bf{\Large The $\bar{s}s$ content of the D-meson at infinite coupling}}\\[1cm]
 C.Alexandrou $^{1,2}$ and   A. Galli $^1$\\[0.5cm]
{\it  $^1$ Paul Scherrer Institute, CH-5232 Villigen, Switzerland\\
$^2$ Department of Natural Sciences, University of Cyprus, CY-1678 Nicosia,
Cyprus\\}
\end{center}

\medskip
\title{}

\vspace*{1cm}

\begin{abstract}
We calculate the $\bar{s}s$ condensate in the D - and B - mesons using the
unquenched hopping parameter expansion at infinite coupling for the Wilson
lattice action. We discuss the
phenomenological relevance of our result.
\end{abstract}

\newpage
\section{Introduction}
Sea quark effects have over the last few year become increasingly more
important in
explaining  a variety of experimental results. Examples are the proton spin
as measured by
the EMC \cite{EMC},
 violations of the Gottfried sum rule obtained by the
NMC \cite{NMC} and confirmed by the NA51 Collaboration \cite{NA51} as well as
abundant $\phi$-meson production in $\bar{p}p$ annihilation \cite{phi} at LEAR.
The experimental indications for sea quark effects are supported by
recent lattice QCD calculations \cite{Japan,liu} of the pion-nucleon
 sigma term where it was  found that the sea contribution is
 twice the valence one
yielding a value for the
sigma term in agreement   with the  result from chiral perturbation
theory \cite{GLS}.
It is therefore reasonable to expect that sea quark contributions
are important in a number of
other observables. In particular they may have implications for
 hadronic decays of the D- and B- mesons.
In ref. \cite{QCD2} it was pointed out that a significant amount of $\bar{s}s$
in the D- and B- mesons would enhance decays of these mesons to strange final
states, thus providing an explanation for the large branching ratios
\cite{fermilab} for $D^0\rightarrow \phi K^0$ and $D^+\rightarrow \phi K^+$.
Similar implications would follow for the decays $B\rightarrow \phi + X$.

In this work we estimate the $\bar{s}s$ content of these mesons using the
unquenched hopping parameter expansion of the Wilson lattice action at infinite
coupling. Infinite coupling corresponds to quark propagation in a background
gluon field. Because of this simplification it is feasible to
perform a
convergent hopping parameter expansion analysis of some
expectation values.
In the quenched approximation one can
define propagators in the random walk representation and then
sum to infinite
order in the hopping parameter \cite{kawamoto1,galli}.
In spite of the fact that the analysis is done at infinite gauge
coupling one still
obtains the meson masses to a few percent
accuracy \cite{kawamoto2}.
In the unquenched case
one can no longer define propagators as random walks and one has to
truncate the expansion to some order and estimate
the systematic errors due to the truncation \cite{hoek}.
In our work we truncate the unquenched expansion at the 8th order.
This still leads to meson masses within 10\% of their experimental values.

The quantity of interest here is the ratio
\be
\frac{<M|\bar{u}u|M>}{<M|\bar{s}s|M>}
\label{ratio}
\ee
where $M$ stands for a D- or a B- meson.
It is shown in the next section  that in the unquenched hopping parameter
expansion   this ratio can be obtained by differentiating the mesonic mass.
We find that the $\bar{s}s$ contribution is suppressed as compared to the
valence
contribution by at least an order of magnitude. We proceed to explain how we
reach this result.

%\section{The Wilson lattice action}
\section{Calculation of mesonic expectation values}
Our starting point is the Wilson lattice action of QCD \cite{wilson}:
We consider an SU(3) color matrix U(b) in the fundamental representation
defined
on each oriented lattice bond b. Our convention is that
\be
U(-b)=U^\dagger(b).
\ee
An oriented path $\om$ on the lattice is a set of bonds
\be
\om=b_1\cup b_2\cup\dots\cup b_n
\ee
such that the end-point of $b_i$ is the starting point of $b_{i+1}$ for
$1\leq i\leq n-1$. We can associate an SU(3) color matrix with $\om$ by
defining
the path ordered product
\be
U(\om)=U(b_1)U(b_2)\dots U(b_n)     \quad.
\ee
The spin matrices are defined in terms of $\gamma$-matrices by
\be
\Gamma(b)=\left\{\begin{array}{cc}
\Gamma^\mu=1+\gamma^\mu&\mbox{if b in $+\mu$ direction}\\
\bar\Gamma^\mu=1-\gamma^\mu&\mbox{if b in $-\mu$ direction}
\end{array}\right.
\ee
where the $\gamma$-matrices are hermitian 4x4 matrices, satisfying
$\{\gamma^\mu,\gamma^\nu\}=2\delta^{\mu\nu}$.\\
The Wilson action is then defined on a lattice $\Lambda$ by
\be
S=-\beta\sum_{p\subset\Lambda^*}TrU(\pa p)-\sum_{f=flavors}\left\{
k_f\sum_{b=\langle xy\rangle}\bar\psi_f(x)
\Gamma(b)U(b)\psi_f(y)-\sum_{x\in\Lambda}
\bar\psi_f(x)\psi_f(x)\right\}
\ee
where $p$ represents a plaquette and the $k_f$ are the hopping parameters
associated with the different quark flavors. $\Lambda^*$ denotes the
set of all bonds defined on $\Lambda$.
The quark fields are represented by the anti-commuting variables
$\psi_f(x)$
and
$\bar\psi_f(x)$ which transform under the $3$ and $\bar 3$ representation of
color.
%\section{Method to calculate mesonic expectation values}

The mesonic expectation value of interest here is then obtained by
evaluating
\bea
&&\langle M^{f_1f_2}(x)|\bar\psi_{f_3}\psi_{f_3}|M^{f_1f_2}(y)\rangle
=\nonumber\\
&&=\frac{1}{Z}\int [d\bar\psi][d\psi][dU]
M^{f_1f_2}(x)^\dagger M^{f_1f_2}(y)
\left(\sum_{z\in\Lambda}\bar\psi_{f_3}(z)\psi_{f_3}(z)\right)
e^{-S(U,\bar\psi,\psi)}\label{exp}
\eea
where $Z$ is the partition function and
$M^{f_1f_2}(x)=\bar\psi_{f_1}(x)M\psi_{f_2}(x)$
represents a meson operator built up of an anti-quark of flavour $f_1$
and a quark of flavour $f_2$ ($M$ is some
matrix in spin and color space).
The three point functions that must be evaluated in order to obtain
the desired ratio are shown in Fig.~1.
They evolve a valence contribution
when $f_3$ is equal to $f_1$ or $f_2$ and
a sea quark contribution.

\begin{figure}
% \vspace{3in}
\centerline{\epsfysize = 3 in \epsffile {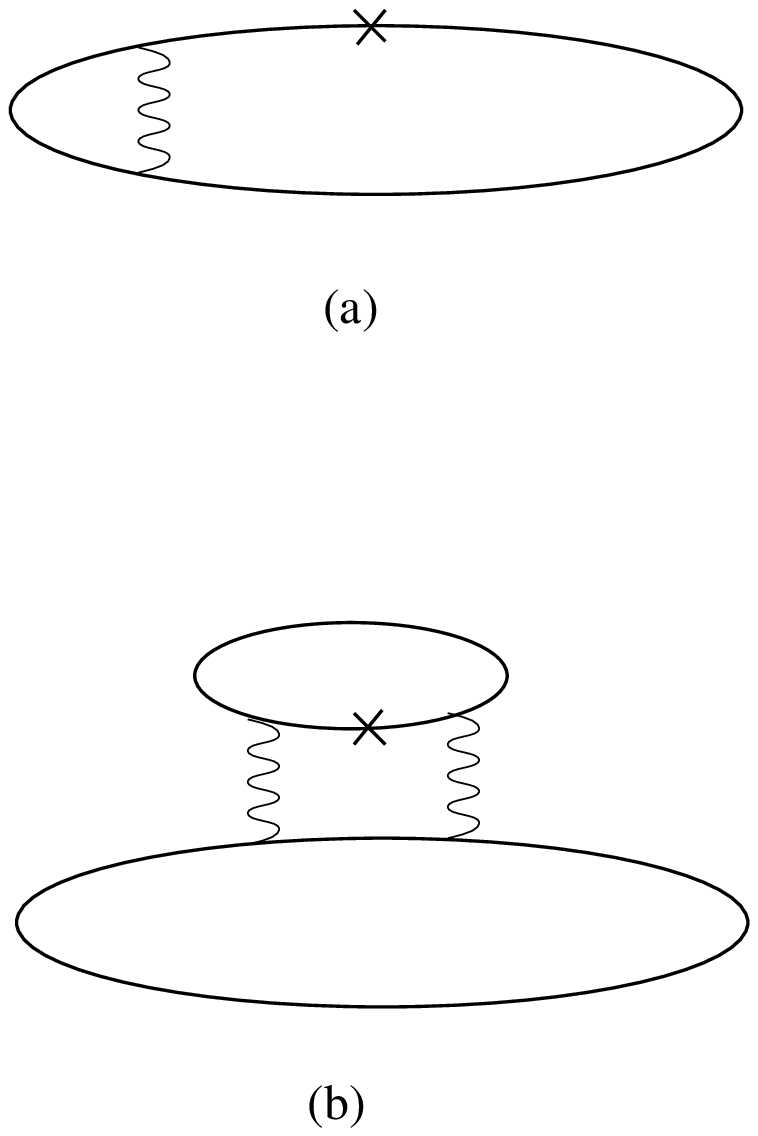}}
\caption{(a) Valence contribution and (b) sea quark contribution to the
ratio given in (1).}
\end{figure}

To avoid the calculation of three point functions one
introduces a parameter
$\lambda_f$ in the action
\be
S_\lambda=-\beta\sum_{p\subset\Lambda^*}TrU(\pa p)-\sum_{f=flavors}\left\{
k_f\sum_{b=\langle xy\rangle}\bar\psi_f(x)
\Gamma(b)U(b)\psi_f(y)-\lambda_f\sum_{x\in\Lambda}
\bar\psi_f(x)\psi_f(x)\right\}   \quad.
\ee
Since
the meson propagator  is given by
\be
\langle M^{f_1f_2}(x)^\dagger M^{f_1f_2}(y)\rangle_\lambda=
\frac{1}{Z_\lambda}\int [d\bar\psi][d\psi][dU]
M^{f_1f_2}(x)^\dagger M^{f_1f_2}(y)e^{-S_\lambda(U,\bar\psi,\psi)}
\label{prop}
\ee
the derivative of the meson propagator
\be
\left.\langle M|\bar\psi_{f_3}\psi_{f_3}|M\rangle _{\rm conn.}=
- \frac{\pa}{\pa\lambda_{f_3}}\langle M^{f_1f_2}(x)^\dagger M^{f_1f_2}(y)
\rangle_\lambda\right|_{\lambda_f=1\forall f}
\ee
leads to the desired expectation value of the connected diagrams in (\ref{exp})
after  setting this parameter to unity to
recover the original action.

The ratio between two different mesonic expectation values can be
expressed as the ratio of the derivative with respect to the
$\lambda$'s of the meson mass, assuming that for large space-time separation
a meson propagator behaves proportionally to
$e^{-m_M|x-y|}$.
\be
\left.\frac{\langle M|\bar\psi_{f}\psi_{f}|M\rangle _{\rm conn.}}
{\langle M|\bar\psi_{f'}\psi_{f'}|M\rangle _{\rm conn.}}=
\frac{\pa m_M(\lambda)/\pa\lambda_f}{
\pa m_M(\lambda)/\pa\lambda_{f'}}\right|_{\lambda_f=1\forall f}
\ee

\section{Hopping parameter expansion for the meson propagator}
Mesonic masses are obtained by studying the long distance behaviour of the
propagators given in (\ref{prop}) . In this work the propagators  are evaluated
using the the hopping parameter expansion at the infinite coupling limit.
This is done by breaking the action $S$ into two
parts, the unperturbed part $S_0$ and the perturbation $S_I$:
\bea
S_0&=&\sum_f\lambda_f\left\{\sum_{x\in\Lambda}\bar\psi_f(x)\psi_f(x)\right\}
\nonumber\\
S_I&=&-\sum_f\left\{
k_f\sum_{b=\langle xy\rangle}\bar\psi_f(x)\Gamma(b)U(b)\psi_f(x)\right\}.
\eea
The plaquette term is not present because we take the infinite coupling limit
i.e. $\beta=0$.
Expanding the exponential in eq. (\ref{prop}) in term of
the perturbation and integrating out the gauge degrees of
freedom $U$ we obtain an expression for the the meson propagator
\be
\langle M^{f_1f_2}(x)^\dagger M^{f_1f_2}(y)\rangle_{\lambda}
=
 \sum_{\omega^+:x\mapsto y;\omega^-:y\mapsto x;\gamma\subset\Lambda^*}C_M~
A(\omega^+\cup\omega^-,\gamma)\tilde k_{f_1}^{|\omega^+|}
\tilde k_{f_2}^{|\omega^-|}\sum_{f_{sea}}\tilde k_{f_{sea}}^{|\gamma|}\label{8}
\ee
where
\be
A(\omega^+\cup\omega^-,\gamma)=
Tr\left(M^\dagger\Gamma(\omega^+)M\Gamma(\omega^-)\right)
{}~Tr\left(\Gamma(\gamma)\right)
{}~ (-1)^{L_\gamma}
{}~\Omega(\omega,\gamma)\label{8p}
\ee
In (\ref{8}) we have rescaled the hopping parameters $k_f$ to
$\tilde k_f=k_f/\lambda_f$.
The term $C_M$ is an overall constant
amplitude which does not affect the meson mass,
$\omega^+$ is a path on the lattice connecting $x$ with $y$,
$\omega^-$ is a path connecting $y$ with $x$,
$\omega=\omega^+\cup\omega^-$ is the combination of them, $\gamma$ is a closed
path representing the sea-quarks loop contributions
and $L_\gamma$ is the number of closed loops formed by the path
$\gamma$. The mapping $\Omega(\omega,\gamma)$ denotes the group
integral over the Haar measure of $SU(3)$ associated with the graph
defined by $\omega$ and $\gamma$.
Finally, $\Gamma(\omega)$ and $U(\omega)$ represent the
ordered products of $\Gamma(b)$ and $U(b)$ on the bonds $b\in\omega$,
respectively.
We do not give a proof of (\ref{8},\ref{8p})
since it is standard \cite{froe,creutz}.

%\section{Meson masses from the hopping parameter expansion}
In order to compute the meson masses we consider the static propagator
\be
G^M(t)=\sum_{\vec x}\langle M((0,\vec 0))^\dagger M((t,\vec x))\rangle
\ee
(we omit the flavour indices for simplicity).
The lowest order diagram representing a static propagator is a double fermion
path $\omega_0$ in the time direction (Fig.2a).
%Since there is a non-zero probability for a
%transition from the lowest order state to a more complicate state
 The full
propagator is given by a sequence of exited states connected by lowest
order
states shown schematically in Fig.2b where each bubble can be viewed as a
process contributing to the excitation. These
excitations renormalize the mass of the static propagator.
The unrenormalized meson mass is given by the
lowest order diagram of the static propagator, namely
\be
G_0^M(t)=C_M\times \exp(-m_0t)=C_M\times(4\tilde k_{f_1}\tilde k_{f_2})^t.
\label{10}
\ee

\begin{figure}
% \vspace{3in}
\centerline{\epsfysize = 8 in \epsffile {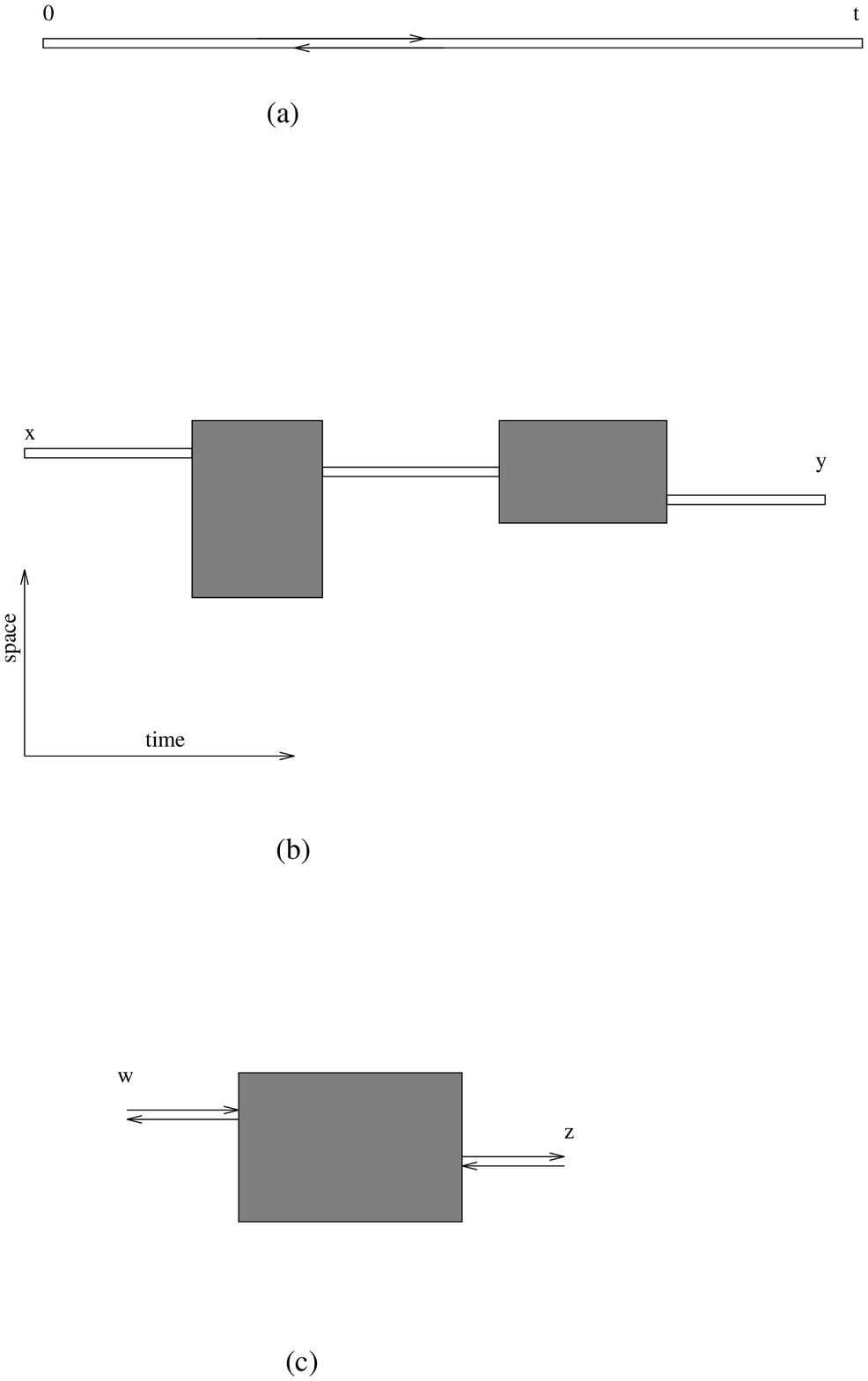}}
\caption{(a) A static unrenormalized meson propagator;
(b) An excitation of the static unrenormalized meson propagator;
(c) A close up of an intermediate excitation.}
\end{figure}

This tells us that the unrenormalized mass is
$m_0=-log(4\tilde k_{f_1}\tilde k_{f_2})$.
Let us now consider the effect on the mass
produced by intermediate exited states. Fig.~2c shows  an excitation,
with the static propagator entering at the initial point $w$
and departing at the final point $z$.
For fixed $w$ and $z$, we  sum over all intermediate exited states to
obtain the total weight for this event. We denote this weight by
$ D^{ M}(w,z)$. The
full propagator takes the form
\bea
\langle M(x)^\dagger M(y)\rangle&=&\sum_{n=0}^{\infty}\sum_{w_i,z_i(i=1\dots
n)}G_0^{{ M}}(x,w_1) D^{ M}(w_1,z_1)G_0^{{ M}}(z_1,w_2)
D^{ M}(w_2,z_2)\times\nonumber\\
& &\times\dots \times  D^{ M}(w_n,z_n)G_0^{{ M}}(z_n,y)
\label{11}
\eea
where the points $w_i$ and $z_i$ are required to be time ordered
and $n$ represents the number of excitations.
This last equation is represented pictorially in Fig.~2b.\\

Following \cite{froe} one can factorize all
static unrenormalized terms $G_0$ in eq (\ref{11}) and compute the
excitations $D(w_i,z_i)$ separately.
We define\footnote{Notice that
$(G_0^{ M}(t))^{-1}=\frac{1}{C^{ M}} e^{m_0t}=\frac{1}{C^{ M}}
(4\tilde k^2)^{-t}$}
\be
\tilde D^{ M}(t)=(G_0^{ M}(t))^{-1}\left.\sum_{x,y}
D^{ M}(x,y)\right|_{y_0-x_0=t}
\ee
where we normalize the excitation relative to the static unrenormalized
propagator over the time interval $t$.
{}From this definition and eq. (\ref{8}) we obtain
\bea
G^{{M}}(t)&=&e^{-m_0t}
\sum_{s_i,t_i(i=1\dots n)}\tilde D^{ M}(t_1-s_1)\dots \tilde D^{ M}(t_n-s_n)=
\nonumber\\
&\simeq&e^{-m_0t}\,e^{p_{{ M}}t}\label{18}
\eea
where $t_i$ and $s_i$ are the corresponding time coordinates of $z_i$ and
$w_i$. The exponential representation of (\ref{18}) is up to an irrelevant
boundary term exact and yields the
 renormalized mass $m_{{ M}}$
\be
m_{{M}}=m_0-p_{{ M}}.
\ee
The mass correction $p_{ M}$ can be written in the form
\be
p_{{M}}=
\sum_{z=(t'\geq 0,\vec z)}(G_0^{ M}(t'))^{-1}D^{ M}(0,z)=
\sum_{z=(t'\geq 0,\vec z)}(4k^2)^{-t'}D^{ M}(0,z) \quad.
\label{pm}
\ee
The last equation follows from (\ref{18}) if we expand the exponential
function containing $p_{ M}$.
 The excitation term
$D^{M}(w,z)$ is defined starting from eq. (\ref{8}) in the following way:
\be
D^{M}(w,z)=\frac{1}{A_0 C_{M}}\sum_{\om:w\mapsto z\mapsto w;\gamma\in
S(\omega)}\,\,
A(\om,\gamma)\tilde k_{f_1}^{|\omega^+|}\tilde
k_{f_2}^{|\omega^-|}\sum_{f_{sea}}
\tilde k_{f_{sea}}^{|\gamma|}
\ee
where the sum is over all closed paths $\om=\om^+\cup\om^-$ from $w$ to $z$
and $S(\omega)\subset\Lambda^*$ is the set of all loops $\gamma$ for with
$\Omega(\omega,\gamma)\neq 0$.
$A^{f_1f_2}(\om,\gamma)$ denotes the amplitude of the intermediate
excitations. $A_0=4$ is a constant which has to be explicitly
evaluated from (\ref{8}).\\

To evaluate the correction to the unrenormalized mass $m_0$ due to the
excitations from eq. (\ref{pm}) we have to identify all closed
paths $\om$ from the point $0$ to
the point $z=(t,\vec z)$ and to sum over all points $\vec z$.
At $\beta=0$ we do not have
plaquettes in the expansion because the plaquette expectation value vanishes,
therefore a diagram consists
of a closed line $\om=\om^+\cup\om^-$
composed by a set of connected bonds coming from the valence-quark
contribution and a second path
$\gamma$ coming from the sea-quark contribution. The group integral
of a diagram is non-vanishing only if, for all $b\in\omega\cup\gamma$,
there exists a second link $b'\in\omega\cup\gamma$ such that $b=-b'$
($b$ is equal $b'$ in opposite direction). A method to easily
evaluate the group integral is discussed in \cite{creutz}.

\section{Results and Conclusions}
We now evaluate the mass of the pseudoscalar and the vector mesons
with $f_1$ and $f_2$ quarks flavours up to order eight in the expansion,
including  valence and sea quarks. The
sea quark contribution arises from  quarks with the same flavour
as the valence quarks, incorporated in the notation $k_{f_1}, k_{f_2}$,
as well as with flavours  different
from the valence quarks  denoted by  $k_{sea}$.
All terms in the expression of the form
$\left ({{\it \tilde k_{f_1}}}^{n}+{{\it \tilde k_{f_2}}}^{n}+{{\it
\tilde k_{sea}}}^{n}\right )
$
with $n=4$ or $n=6$ correspond to sea quark contributions to the meson
mass.
Collecting all terms we obtain for the pseudoscalar meson mass

\bea
m_{PS}(\{\tilde k_{f_1}\tilde k_{f_2}\};\tilde k_{sea})=\\
&&-\ln (4\,{\it \tilde k_{f_1}}\,{\it \tilde k_{f_2}})-
3\,{\it \tilde k_{f_1}}\,{\it \tilde k_{f_2}}
-36\,{{\it \tilde k_{f_1}}}^{2}{
{\it \tilde k_{f_2}}}^{2}-
432\,{{\it \tilde k_{f_1}}}^{3}{{\it \tilde k_{f_2}}}^{3}
\nonumber\\&&
-3\,{{\it \tilde k_{f_2}}}^{2}
\left ({{\it \tilde k_{f_1}}}^{4}+{{\it \tilde k_{f_2}}}^{4}+{{\it
\tilde k_{sea}}}^{4}\right )
\nonumber\\&&
-19\,{{\it
\tilde k_{f_1}}}^{2}\left ({{\it \tilde k_{f_1}}}^{4}+{{\it \tilde
k_{f_2}}}^{4}+{{\it \tilde k_{sea}}}^{4}\right )-{
\it \tilde k_{f_1}}\,{\it \tilde k_{f_2}}\,\left ({{\it \tilde
k_{f_1}}}^{4}+{{\it \tilde k_{f_2}}}^{4}+{{\it \tilde k_{sea}}}^{4}
\right )
\nonumber\\&&
-5328\,{{\it \tilde k_{f_1}}}^{4}{{\it \tilde k_{f_2}}}^{4}-8\,{{\it \tilde
k_{f_1}}}^{2}\left ({{
\it \tilde k_{f_1}}}^{6}+{{\it \tilde k_{f_2}}}^{6}+{{\it \tilde
k_{sea}}}^{6}\right )-8\,{{\it \tilde k_{f_2}}}^{2}
\left ({{\it \tilde k_{f_1}}}^{6}+{{\it \tilde k_{f_2}}}^{6}+{{\it
\tilde k_{sea}}}^{6}\right )
\nonumber\\&&
-16\,{\it
\tilde k_{f_1}}\,{\it \tilde k_{f_2}}\,\left ({{\it \tilde k_{f_1}}}^{6}+{{\it
\tilde k_{f_2}}}^{6}+{{\it \tilde k_{sea}}}^{6}
\right )-32\,{\it \tilde k_{f_1}}\,{{\it \tilde k_{f_2}}}^{3}\left ({{\it
\tilde k_{f_1}}}^{4}+{{\it \tilde k_{f_2}}}^
{4}+{{\it \tilde k_{sea}}}^{4}\right )
\nonumber\\&&
-32\,{{\it \tilde k_{f_1}}}^{3}{\it \tilde k_{f_2}}\,\left ({{\it \tilde
k_{f_1}
}}^{4}+{{\it \tilde k_{f_2}}}^{4}+{{\it \tilde k_{sea}}}^{4}\right )-48\,{{\it
\tilde k_{f_1}}}^{2}{{\it
\tilde k_{f_2}}}^{2}\left ({{\it \tilde k_{f_1}}}^{4}+{{\it \tilde
k_{f_2}}}^{4}
+{{\it \tilde k_{sea}}}^{4}\right )
\nonumber\\&&
+8
\,{{\it \tilde k_{f_2}}}^{4}\left ({{\it \tilde k_{f_1}}}^{4}+{{\it \tilde
k_{f_2}}}^{4}+{{\it \tilde k_{sea}}}^{4}
\right )+8\,{{\it \tilde k_{f_1}}}^{4}\left ({{\it \tilde k_{f_1}}}^{4}+{{\it
\tilde k_{f_2}}}^{4}+{{\it
\tilde k_{sea}}}^{4}\right )\nonumber
\eea
and for the vector meson mass
\bea
m_{VK}(\{\tilde k_{f_1}\tilde k_{f_2}\};\tilde k_{sea})=\\&&
-\ln (4\,{\it \tilde k_{f_1}}\,{\it \tilde k_{f_2}})-2\,{\it \tilde
k_{f_1}}\,{\it \tilde k_{f_2}}-16\,{{\it \tilde k_{f_1}}}^{2}{
{\it \tilde k_{f_2}}}^{2}-128\,{{\it \tilde k_{f_1}}}^{3}
{{\it \tilde k_{f_2}}}^{3}
\nonumber\\&&
-3\,{{\it \tilde k_{f_2}}}^{2}
\left ({{\it \tilde k_{f_1}}}^{4}+{{\it \tilde k_{f_2}}}^{4}
+{{\it \tilde k_{sea}}}^{4}\right )
\nonumber\\&&
-19\,{{\it
\tilde k_{f_1}}}^{2}\left ({{\it \tilde k_{f_1}}}^{4}+{{\it \tilde
k_{f_2}}}^{4}+{{\it \tilde k_{sea}}}^{4}\right )-{
\it \tilde k_{f_1}}\,{\it \tilde k_{f_2}}\,\left ({{\it \tilde
k_{f_1}}}^{4}+{{\it \tilde k_{f_2}}}^{4}+{{\it \tilde k_{sea}}}^{4}
\right )
\nonumber\\&&
-1088\,{{\it \tilde k_{f_1}}}^{4}{{\it \tilde k_{f_2}}}^{4}-8\,{{\it \tilde
k_{f_1}}}^{2}\left ({{
\it \tilde k_{f_1}}}^{6}+{{\it \tilde k_{f_2}}}^{6}+{{\it \tilde
k_{sea}}}^{6}\right )-8\,{{\it \tilde k_{f_2}}}^{2}
\left ({{\it \tilde k_{f_1}}}^{6}+{{\it \tilde k_{f_2}}}^{6}+{{\it \tilde
k_{sea}}}^{6}\right )
\nonumber\\&&
-16\,{\it
\tilde k_{f_1}}\,{\it \tilde k_{f_2}}\,\left ({{\it \tilde k_{f_1}}}^{6}+{{\it
\tilde k_{f_2}}}^{6}+{{\it \tilde k_{sea}}}^{6}
\right )-24\,{\it \tilde k_{f_1}}\,{{\it \tilde k_{f_2}}}^{3}\left ({{\it
\tilde k_{f_1}}}^{4}+{{\it \tilde k_{f_2}}}^
{4}+{{\it \tilde k_{sea}}}^{4}\right )
\nonumber\\&&
-24\,{{\it \tilde k_{f_1}}}^{3}{\it \tilde k_{f_2}}\,\left ({{\it \tilde
k_{f_1}
}}^{4}+{{\it \tilde k_{f_2}}}^{4}+{{\it \tilde k_{sea}}}^{4}\right )-32\,{{\it
\tilde k_{f_1}}}^{2}{{\it
\tilde k_{f_2}}}^{2}\left ({{\it \tilde k_{f_1}}}^{4}+{{\it \tilde
k_{f_2}}}^{4}+{{\it \tilde k_{sea}}}^{4}\right )
\nonumber\\&&
+8
\,{{\it \tilde k_{f_2}}}^{4}\left ({{\it \tilde k_{f_1}}}^{4}+{{\it \tilde
k_{f_2}}}^{4}+{{\it \tilde k_{sea}}}^{4}
\right )+8\,{{\it \tilde k_{f_1}}}^{4}\left ({{\it
\tilde k_{f_1}}}^{4}+{{\it \tilde k_{f_2}}}^{4}+{{\it
\tilde k_{sea}}}^{4}\right )\nonumber
\eea

%\section{Fixing the parameters and checking the expansion}

To fix the quark masses  we use the experimental values of
meson masses expressed in units of the $\rho$ mass. For the u- ,
d- and s- quark masses we use the experimental pion and kaon masses
(we consider the u- and
the d- quarks to have the same  mass).
For this evaluation we neglect all sea quarks of
flavour different from u, d and s.
We thus obtain two equations
\be
\frac{m_{PS}(\{k_u=k_d\};k_{sea}=k_s)}{
m_{VK}(\{k_u=k_d\};k_{sea}=k_s)}=\frac{m_\pi}{m_\rho}\simeq 0.18
\ee
\be
\frac{m_{PS}(\{k_u,k_s\};k_{sea}=0)}{
m_{VK}(\{k_u=k_d\};k_{sea}=k_s)}=\frac{m_K}{m_\rho}\simeq 0.64
\ee
from which we can extract the values of
$k_u$ and $k_s$ .
The resulting values are  $k_u=k_d= 0.28(3)$ and  $k_s=0.25(2)$.
The error given here is an estimate of the systematic error from the
truncation of the expansion. It was obtained by comparing
the truncated results to the known bounds
in the meson masses \cite{galli}
\footnote{ It was shown by the same
author that a very good  approximation to the meson mass is given by
$m_{PS}\simeq -\log\left[4k_{f_1}k_{f_2}/(1-12k_{f_1}k_{f_2})\right]$ and
$m_{VK} \simeq -\log\left[4k_{f_1}k_{f_2}/(1-8k_{f_1}k_{f_2})\right]$. The
deviation of the
truncated results from these values can also be used  to estimate   the
systematic error due to the truncation. One obtains a similar value as
the one quoted in the main text.}
\be
m_{PS} \leq -\log(4k^2) -\frac{3k^2}{1-12k^2} \hspace*{1cm}
m_{VK} \leq -\log(4k^2) -\frac{2k^2}{1-8k^2} \hspace*{1cm}
\ee
where $k={\rm max}(k_{f_1},k_{f_2},k_{sea})$.

Using these parameters one can check the expansion by predicting
the $K^*$ mass obtained from the vector meson mass formula
\be
\frac{m_{VK}(\{k_u=0.28,k_s=0.25\};k_{sea}=0)}{
m_{VK}(\{k_u=k_d=0.28\};k_{sea}=k_s)}=1.25(12)
\ee
which compares favourably with the experimental value of
$m_{K^*}/m_\rho \simeq1.16$. Knowing the u- and s- quark masses one can
estimate  the ratio of the $\bar{s} s $ to $\bar{u} u $
condensates in the vaccum. To order eight it is easy to compute and
we find $<\bar{s} s>/<\bar{u} u>=\left( k_s/k_u \right )^8 $. Because
this depends on the eighth power of the $k-$ values it is very sensitive on
the errors of $k_u$ and $k_s$.

In an analogous way  one fixes the c-quark mass
using the experimental value of the $D$- meson mass. The accuracy
of the expansion can then be checked by comparing the predicted mass of
the $D^*$ with the experimental value. In this case we find
$k_c=0.14(1)$ leading to a
 $D^*$ meson  mass ratio $m_{D^*}/m_\rho$ of 2.5(1)
 to be compared with the experimental value of 2.62.

\noindent
Having fixed the quark masses we obtain for the ratio

\be
\left.\frac{\langle D|\bar u u|D\rangle}{\langle D|\bar s s|D\rangle}=
\frac{\pa m_{PS}(\tilde k_u;k_c=0.14;k_{sea}=k_s)/\pa\lambda_u}
{\pa m_{PS}(k_u=0.28;k_c=0.14;\tilde k_s)/\pa\lambda_s}
\right|_{\tilde k_u=0.28;\tilde k_s=0.25} \simeq 50(10)\label{ratio1}
\ee
The error given for the ratio is estimated by taking into account
 only the errors in the values of
the $k$'s which include part of the systematic error due to the truncation. The
latter can not
be estimated in the same way as the error in the $k$'s because no
upper bounds exist for the derivative of the meson mass.
%We can
%still, however, give a lower bound for the ratio as done below.

The same analysis can also be done in the case of the B- meson where one
uses the experimental value of the $B$- meson mass to fix $k_b$.
We find
$k_b=0.011(2)$ leading to a
 $B^*$ meson  mass ratio $m_{B^*}/m_\rho$ of 6.8(4)
 within  the experimental value of 7.06 and for the ratio
\be
\frac{\langle B|\bar u u|B\rangle}{\langle B|\bar s s|B\rangle}
\simeq 45(5) \label{ratio2}\quad.
\ee

We note that, for
the  allowed parameter range
$(k_u,k_f,k_s)\in[0,0.28]^3$ where $k_f$ is the hopping parameter
corresponding to the c- or the b- quark
the ratios (\ref{ratio1},\ref{ratio2}) remain bounded
from below by
\be
\frac{\langle M|\bar u u|M\rangle}{\langle M|\bar s s|M\rangle}
>40~~~~\forall (k_u,k_f,k_s)\in[0,0.28]^3             \quad.
\ee

{}From the above values of the ratios  we conclude that the non-valence
contribution
in the D- and B-  mesons is
suppressed compared to the valence contribution by at least an order of
magnitude.
This would comply with the OZI
rule meaning that the OZI  rule can not be evaded in the limit of infinite
coupling.
Therefore the enhancement observed in D- meson decays to strange final states
seems not to be explained by a rich $\bar s s$ content in the D-meson.
Whether  finite $\beta$ effects can change the above ratio by an order of
magnitude can only be checked by doing a finite $\beta$ lattice simulation.\\
\\
{\bf Acknowledgements:} We thank M. Karliner for the initial motivation
for this work and interesting comments as well as F. Jegerlehner for a
careful reading of the manuscript.

\end{document}